\documentstyle[pre,aps]{revtex}

\newcommand{\be}{\begin{equation}}
\newcommand{\ee}{\end{equation}}
\newcommand{\sR}{\stackrel{\rightarrow}{R}}
\newcommand{\sS}{\stackrel{\rightarrow}{S}}
\newcommand{\sB}{\stackrel{\rightarrow}{B}}
\newcommand{\sse}{\stackrel{\rightarrow}{e}}
\newcommand{\sh}{\stackrel{\rightarrow}{h}}
\newcommand{\sr}{\stackrel{\rightarrow}{r}}
\newcommand{\ssb}{\stackrel{\rightarrow}{b}}
\newcommand{\ra}{\rightarrow}
\newcommand{\al}{\alpha}
\newcommand{\bt}{\beta}
\newcommand{\om}{\omega}
\newcommand{\gm}{\gamma}
\newcommand{\vp}{\varphi}
\newcommand{\ep}{\varepsilon}
\newcommand{\lbd}{\lambda}
\newcommand{\prt}{\partial}

\begin{document}

\draft

\title{ Collimation Mechanism for Atom Lasers\footnote{Invited report,
International Workshop on Laser Physics, Berlin, July 1998}} 

\author{V.I. Yukalov$^{1,2}$ and E.P. Yukalova$^{3,4}$ }

\address{$^1$Centre for Interdisciplinary Studies in Chemical Physics \\
University of Western Ontario, London, Ontario N6A 3K7, Canada \\ [3mm]
$^2$Bogolubov Laboratory of Theoretical Physics \\
Joint Institute for Nuclear Research, Dubna 141980, Russia \\ [3mm]
$^3$ Department of Physics and Astronomy \\
University of Western Ontario, London, Ontario N6A 3K7, Canada \\ [3mm]
$^4$Laboratory of Computing Techniques and Automation \\
Joint Institute for Nuclear Research, Dubna 141980, Russia}

\maketitle

\begin{abstract}

A mechanism is suggested for creating well--collimated beams of neutral
spin--polarized particles by means of magnetic fields. This mechanism can
be used in atom lasers for the formation of directed coherent beams of
atoms. The directed motion of atoms is achieved only with the help of
magnetic fields, no mechanical collimators being involved.

\end{abstract}

\newpage

\section{Introduction}

A device emitting a coherent atomic beam, similar to a laser radiating
coherent photon rays, is called atom laser [1--7]. Bose atoms can be
prepared in the Bose--condensed state by cooling them in a trap [8--10].
The Bose--condensed state is believed to be a coherent state [11], which
is supported by the interference experiments [12,13].

It is worth noting that a Bose--condensed state cannot be a pure coherent
state, but it can be only partially coherent. This is because a coherent
state is not an eigenstate of a Bose system [14]. Another problem is that,
if one identifies the Bose--condensed state with a state having broken
gauge symmetry, then there appear anomalous fluctuations of the number of
particles, with nonthermodynamic behaviour [15,16]. This anomalous
behaviour originates from infrared divergences typical of any Bose fluids
with broken gauge symmetry [17--19] and is related to the fluctuations of
phases of field operators [20].

However, a system of finite number of particles in a trap does not undergo
a genuine phase transition with breaking symmetry [21--24], similarly to
the case of the Bose gas in restricted geometries [25]. In such cases, the
singularities of all thermodynamic functions are rounded off and the latter 
become analytic, varying smoothly. This means that Bose condensation of a 
finite number of particles in a trap is a crossover phenomenon that is not
accompanied by breaking gauge symmetry. Although breaking the latter can
be employed as a technical trick facilitating some calculations, for
instance, for considering collective excitations of a trapped Bose gas
[26--29]. The usage of this trick is limited by the cases when it does not
lead to anomalies, like those for the fluctuations of the number of 
particles. The Bose condensation in a trap, being a crossover phenomenon,
occurs at a crossover temperature that can be defined as a temperature at
which at least one of thermodynamic characteristics has a maximum. This
definition of a crossover temperature is typical of crossover phenomena
[21--25,30,31].

Assume that Bose atoms are trapped and cooled down to experience the Bose
condensation, when, at least partly, they become coherent. An important
problem in realizing an atom laser is how to form a directional beam of
atoms. To make the output highly directional is, actually, the first
condition on a laser [7]. However, at the present time there are no
mechanisms permitting one to create such well collimated beams of neutral
atoms flying out of traps. To form an output coupler, one employs short
radio--frequency pulses transferring atoms from trapped states to an
untrapped state. But, when escaping from a trap, atoms fly out more or
less in all directions [12,13,32], with an inisotropy formed by the
gravitational force. This makes the principal difference between the
present--day traps and optical lasers. "The major difference between the
two devices is that the photons from an optical laser generally emerge in
a well--collimated beam, whereas atoms from the atom laser fly out in all
directions" [33]. In order to force the escaping atoms to move in one
preferable direction, one can employ some mechanical blocking or external
laser beams (see discussion in Ref. [34]). For example, to provide a
directed output beam, an implementation of the laser scheme using hollow
optical fibers was suggested [35], with a momentum kick provided by two
laser beams.

In this paper, we consider a new mechanism for creating well--collimated
beams of neutral atoms. This mechanism does not require any mechanical
blocking or additional laser operation. Collimation of an atomic beam is
achieved only by means of magnetic fields of a trap. A particular
configuration of the trap magnetic field, leading to a semiconfining
regime of motion, has been studied earlier [36--38]. Here we generalize
the consideration showing that there exists a large class of magnetic
fields permitting one to create directed beams of atoms for atom lasers.

\section{Evolution Equations}

When the space variation of magnetic fields is sufficiently smooth, the
dynamics of neutral atoms is well described in the semiclassical
approximation [36,37] for the quantum--mechanical averages of the
real--space coordinate, $\sR\equiv<\hat{\sR}>=\{ R_\al\}$, and of the spin
operator, $\sS\equiv<\hat{\sS}>=\{ S_\al\}$, where $\al=x,y,z$. Then, for
an atom of mass $m$ and magnetic moment $\mu_0$, the equation for the
average space variable is
\be
\frac{d^2R_\al}{dt^2} = \frac{\mu_0}{m}\sS\cdot\frac{\prt\sB}{\prt R_\al}
\ee
and the evolution of the average spin is given by the equation
\be
\frac{d\sS}{dt} = \frac{\mu_0}{\hbar}\sS\times \sB \; .
\ee
The total magnetic field
\be
\sB = \sB_1 + \sB_2
\ee
consists of two terms. One is the quadrupole field
\be
\sB_1 = B_1'\left ( R_x\sse_x + R_y\sse_y + \lbd R_z\sse_z\right ) \; ,
\ee
in which $\sse_\al$ are the unit Cartezian vectors and $\lbd$ is an
anisotropy parameter. If one requires that $\sB_1$ satisfies the equation
$\stackrel{\ra}{\nabla}\cdot\sB_1=0$, then $\lbd=-2$. But if the field (4)
is produced by several different magnetic coils, then one can, in general,
realize any anisotropy with an arbitrary parameter $\lbd$. The second term
in (3) is a transverse field
\be
\sB_2 = B_2\sh(t)\; , \qquad \sh(t) = h_x\sse_x + h_y\sse_y \; ,
\ee
in which $h_\al=h_\al(t)$ and
$$
|\sh|^2 = h_x^2 + h_y^2 = 1 \; , \qquad \sh =\sh(t) \; .
$$

It is convenient to work with the dimensionless space variable
\be
\sr \equiv\frac{\sR}{R_0} = \{ x,y,z\} \; , \qquad
R_0\equiv \frac{B_2}{B_1'} \; .
\ee
Introduce also the characteristic frequencies
\be
\om_1^2 \equiv\frac{\mu_0 B_1'}{mR_0} \; , \qquad
\om_2 \equiv\frac{\mu_0 B_2}{\hbar} \; .
\ee
Then equation (1) can be written as
\be
\frac{d^2\sr}{dt^2} = \om_1^2\left ( S_x\sse_x + S_y\sse_y +
\lbd S_z\sse_z\right ) \; .
\ee
This equation is to be complemented by initial conditions
\be
\sr(0) =\sr_0 = \{ x_0,y_0,z_0\} \; , \qquad \dot{\sr}(0) =\dot{\sr}_0 =
\{\dot{x}_0,\dot{y}_0,\dot{z}_0\} \; ,
\ee
in which the dot means the time derivative.

The equation (2) for the average spin can be reduced to the form
\be
\frac{d\sS}{dt} = \om_2\hat A\sS \; ,
\ee
where the matrix $\hat A=[A_{\al\bt}]$ consistes of the elements
$$
A_{\al\bt} = - A_{\bt\al}\; , \qquad A_{\al\al} = 0 \qquad 
(\al,\bt=1,2,3) \; ,
$$
\be
A_{12} =\lbd z \; , \qquad A_{23} = x + h_x\; , \qquad 
A_{31} = y + h_y \; .
\ee
The initial condition for (10) is
\be
\sS(0) =\sS_0 = \{ S_x^0,S_y^0,S_z^0\} \; .
\ee

Assume that the following inequalities hold true:
\be
|\om_1|\ll \om_2 \; , \qquad \left |\frac{d\sh}{dt}\right | \ll \om_2\; .
\ee
Defining an effective frequency $\om=\om(t)$ of the transverse field,
\be
\om\equiv\left | \frac{d\sh}{dt}\right |\; ,
\ee
we may rewrite the inequalities (13) as
\be
\left | \frac{\om_1}{\om_2}\right | \ll 1 \; , \qquad
\left | \frac{\om}{\om_2}\right | \ll 1 \; .
\ee
The existence of these inequalities makes it possible to find approximate
solutions to Eqs. (8) and (10) by employing the scale separation approach
[39--41].

\section{Scale Separation}

The first step in the scale separation approach [39--41] is to classify
the functions under consideration into fast and slow. Inequalities (15)
suggest that the variables $\sr$ and $\sh$ are slow, as compared to the
fast variables $\sS$. It is worth noting that equation (8) for $\sr$ is a
differential equation of second order, while that for $\sS$ is a
first--order differential equation. Hence, the set of these differential
equations does not make what is called the standard form of a dynamical
system [42], where all equations are to be first--order differential
equations. Then one may ask whether it is admissible to treat the
variable $\sr$ as slow with respect to $\sS$, when not their first
derivatives are compared. However, it is not difficult to show, by a
simple change of notations, as is done in Appendix A, that it is really
the case: the variable $\sr$ is low as compared to $\sS$.

With the slow variables treated as quasi--integrals of motion, the
solution of Eq. (10) can be found in the following way. Solve the
eigenproblem
\be
\hat A\ssb_i =\al_i\ssb_i \; ,
\ee
resulting in the eigenvalues
$$
\al_{1,2} = \pm i\al \; , \qquad \al_3 = 0\; , \qquad
\al^2\equiv A_{12}^2 + A_{23}^2 + A_{31}^2 \; ,
$$
and in the eigenvectors
$$
\ssb_i =\frac{1}{\sqrt{C_i}}\left [ \left ( A_{12}A_{23} - \al_i A_{31}
\right ) \sse_x + \left ( A_{12}A_{31} + \al_i A_{23}\right ) \sse_y
+ \left ( A_{12}^2 + \al_i^2\right ) \sse_z\right ] \; ,
$$
with the normalization constant $C_i$ given by the equation
$$
C_i = \left ( A_{12}^2 - |\al_i|^2\right )^2 +
\left ( A_{12}^2 + |\al_i|^2\right )
\left ( A_{23}^2 + A_{31}^2\right ) \; .
$$
It is possible to check that the eigenvectors $\ssb_i$ satisfy the
orthonormality condition
\be
\ssb_i^*\cdot\ssb_j =\delta_{ij} \; .
\ee
When the transverse field $\sh$ is kept constant, then a particular
solution of (10) has the form of $\ssb_i\exp(\al_i\om_2 t)$, and the
general solution is
$$
\sS^{(0)}(t) = \sum_{i=1}^3 a_i\ssb_i\exp(\al_i\om_2t) \qquad
(\sh = const)\; .
$$
Substituting here the dependence of $\sh=\sh(t)$ on time and using the
notation
\be
\ssb_i(t) \equiv \ssb_i\left ( \sh(t)\right ) \; , \qquad
\al_i(t) \equiv \al_i\left (\sh(t)\right ) \; ,
\ee
we obtain an approximate solution of Eq. (10) in the form
\be
\sS(t) = \sum_{i=1}^3 a_i\ssb_i(t)\exp\left\{ \al_i(t)\om_2 t \right\} \; .
\ee
The coefficients $a_i$ can be defined from the initial condition (12),
with the use of the orthonormality condition (17), which gives
\be
a_i = \sS_0\cdot\ssb_i(0) \; .
\ee

A slightly different presentation of an approximate solution to Eq. (10)
can be found as follows. Let us look for a particular solution of the type
$\ssb_i\exp(\vp_i)$, keeping in mind that $\ssb_i$ and $\vp_i$ can depend
on time. Substituting this particular solution into (10) yields
\be
\ssb_i\dot{\vp}_i + \dot{\ssb}_i = \om_2\al_i\ssb_i \; ,
\ee
where the dot means the time derivative. Multiplying (21) by $\ssb_i^*$,
we get
$$
\dot{\vp}_i = \om_2\al_i - \ssb_i^*\cdot\dot{\ssb}_i \; ,
$$
from where
\be
\vp(t) = 
\int_0^t\left (\om_2\al_i - \ssb_i^*\cdot\dot{\ssb}_i\right ) dt \; .
\ee
Substituting (22) back into (21) gives the equation
\be
\dot{\ssb}_i =\left ( \ssb_i^*\cdot\dot{\ssb}_i\right ) \ssb_i \; ,
\ee
playing the role of the criterion for the validity of (22). If (23) is
valid, then the solution to (10) writes
\be
\sS(t) = \sum_{i=1}^3 a_i\ssb_i(t)\exp\{\vp_i(t)\} \; .
\ee

Condition (23) imposes a restriction on the behaviour of $\ssb_i$ and it
may be not valid exactly for an arbitrary field $\sh(t)$. However, here we
have the parameter $\ep\equiv|\om/\om_2|$, which, accroding to (15), is
small, $\ep\ll 1$. As far as $\ssb_i$ depends on time through $\sh(t)$,
then $\dot{\ssb}_i\ra 0$ when $\ep\ra 0$. Therefore, equation (23) becomes
asymptotically exact when $\ep\ra 0$. And for small $\ep\ll 1$, equation
(23) is asymptotically valid for an arbitrary field $\sh(t)$ satisfying
(15). Thus, the asymptotic form (24) is an approximate solution to
equation (10). With the small parameter $\ep\ll 1$, the phase (22) is
asymptotically equal to
$$
\vp(t)\simeq \om_2\al_i(t)t \qquad (\ep\ll 1)\; ,
$$
which reduces solution (24) to the form (19).

In this way, expressions (19) and (24) are both approximate solutions to
equation (10), provided that the inequality $\ep\ll 1$ is valid. When
$\ep\ra 0$, these solutions tend to the exact solution corresponding to
$\ep=0$. Because of this, such solutions are often called asymptotically
exact with respect to $\ep\ra 0$.

The found solution for the fast variable $\sS$ has to be substituted
into the equation (8) for the slow variable $\sr$, averaging the
right--hand side of this equation over time according to the rule
\be
< f > \equiv \lim_{\tau\ra\infty}\frac{1}{\tau}\int_0^\tau
f(\sr,t) dt \; ,
\ee
with the slow variable $\sr$ kept fixed. This averaging procedure
transforms (8) to the equation
\be
\frac{d^2\sr}{dt^2} =  \stackrel{\ra}{F} \; ,
\ee
with the averaged force
\be
\stackrel{\ra}{F} \equiv \om_1^2 < S_x\sse_x + S_y\sse_y +
\lbd S_z\sse_z > \; .
\ee
Note that any form of $\sS$, either (19) or (24), can be used, since in
both these cases we come to the equality
\be
< \sS > = a_3 < \ssb_3 > \; ,
\ee
in which
\be
a_3 = \frac{(x + h_x^0)S_x^0 +(y + h_y^0)S_y^0 +\lbd z S_z^0}
{[(x+h_x^0)^2 + (y+h_y^0)^2 +\lbd^2z^2]^{1/2}} \; , \qquad
b_3 = \frac{(x + h_x)\sse_x +(y + h_y)\sse_y +\lbd z\sse_z}
{[(x+h_x)^2 + (y+h_y)^2 +\lbd^2z^2]^{1/2}} \; ,
\ee
with $h_\al^0\equiv h_\al(0)$ and $\al=x,y,z$. In this way, for the
averaged force (27) we have
\be
\stackrel{\ra}{F} = \om_1^2 a_3 < b_3^x\sse_x + b_3^y\sse_y
+ \lbd b_3^z\sse_z > \; ,
\ee
where $b_3^\al$, with $\al=x,y,z$, are the components of the vector
$\ssb_3=\{ b_3^\al\}$.

\section{Rotating Field}

To proceed further, let us concretize the transverse field $\sh(t)$ in
(5). Take this field in the form of the rotating field with
\be
h_x =\cos\om t \; , \qquad h_y =\sin\om t\; ,
\ee
which is employed in some magnetic traps [43,44]. Then, using $h_x^0=1,\;
h_y^0=0$, and the equality 
$$
< x\cos\om t + y\sin\om t > = 0 \; ,
$$
for the force (30) we find
\be
\stackrel{\ra}{F} =\frac{\om_1^2[(1+x)S_x^0+yS_y^0+\lbd zS_z^0)]
(x\sse_x+y\sse_y+2\lbd^2 z\sse_z)}{2[(1+2x+x^2+y^2+\lbd^2z^2)
(1+x^2+y^2+\lbd^2z^2)]^{1/2}}\; .
\ee

If we take for the initial spin polarization the standard initial
condition $S_x^0\neq 0,\; S_y^0=S_z^0=0$ that is used for confining atoms
in a trap, then, for $S_x^0<0$, the force (32) really provides
confinement, when atoms oscillate in a nearly harmonic potential. This
type of the confined oscillating motion does not change much if we add to
(32) the gravitational force. A vertical field gradient supplies the
levitating force to support the atom against gravity. The combination of
the magnetic field and gravity produces a very nearly harmonic confining
potential within the trap volume in all three dimensions, so that the
atom oscillates around an effective equilibrium position [45,46]. The role
of the gravitational force is discussed in Appendix B.

We take here a different type of the initial spin polarization
corresponding to the initial condition
\be
S_x^0 = S_y^0 = 0\; , \qquad S_z^0\equiv S\neq 0\; .
\ee
The choice of initial conditions, as is well known from quantum mechanics,
is not prescribed a priori but can be realized in any desirable way. Some
more details on the possibility of realizing different initial conditions
for the spin polarization are considered in Appendix C.

To simplify the evolution equations, it is convenient to measure time in
units of $\om_0^{-1}$, where
\be
\om_0^2 \equiv |S\om_1^2| \; .
\ee
In what follows, we shall deal with this dimensionless time. To return to
the dimensional time, we need to set $t\ra\om_0t$.

In order to take into account the finite size of a trap, we introduce the
shape factor
\be
\vp(\sr) =\exp\left\{ -\frac{1}{L^2}\left ( x^2 + y^2 +\ep^2 z^2\right )
\right \} \; ,
\ee
in which $L$ is the characteristic radius of the device in the radial
direction and $L/\ep$ is the length of the device in the axial direction.
And let us define the function
\be
f(\sr)= 
\frac{\vp(\sr)}{[(1+2x+x^2+y^2+\lbd^2z^2)(1+x^2+y^2+\lbd^2z^2)]^{1/2}}\; .
\ee

With these definitions, equation (26), for the initial spin polarization
(33), results in the system of equations for the components
\be
\frac{d^2x}{dt^2} = \frac{\lbd}{2} \left ({\rm sgn} S\right ) fxz \; ,
\qquad
\frac{d^2y}{dt^2} = \frac{\lbd}{2} \left ({\rm sgn} S\right ) fyz \; ,
\qquad
\frac{d^2z}{dt^2} = \lbd^3 \left ({\rm sgn} S\right ) fz^2 \; .
\ee
These equations are invariant with respect to the transformations
$\lbd\ra -\lbd,\; S\ra -S$ as well as to the transformations $S\ra -S,\;
z\ra -z$. Therefore, it is sufficient to study only one case for which the
sign of $\lbd S$ is fixed, since the change $\lbd S\ra -\lbd S$ leads to a
picture that is mirror symmetric with respect to the $x-y$ plane. Take,
for instance,
\be
\lbd S > 0 \; , \qquad |\lbd|\equiv\bt \; .
\ee
Changing $\lbd S$ by $-\lbd S$ would invert the whole picture according to
the tranformation $z\ra -z$. We may also notice that the system of
equations (37) possesses the integral of motion
\be
xy\frac{d}{dt}\ln\frac{y}{x} = const \; ,
\ee
which shows that the motion along the $x$ and $y$ axes are similar to each
other, so that, under the same initial conditions for $x$ and $y$, the
laws of motion $x(t)$ and $y(t)$ coincide. Because of this, we consider in
what follows only the $x$ component. Thus, taking account of (38), from
(37) we have
\be
\frac{d^2x}{dt^2} =\frac{\bt}{2} f xz \; , \qquad
\frac{d^2z}{dt^2} =\bt^3 fz^2 \; .
\ee

At the initial stage of the process, when $|\sr|\ll 1$ and $f(\sr)\simeq
1$, the equations (40) can be solved analytically. Then the second
equation from (40) writes
\be
\frac{d^2z}{dt^2} =\bt^3 z^2 \; .
\ee
This can be integrated once yielding
\be
\left (\frac{dz}{dt}\right )^2 = \frac{2}{3}\bt^3\left ( z^3 - z_0^3
+\zeta\right ) \; ,
\ee
where
\be
\zeta\equiv\frac{3}{2\bt^3}\dot{z}_0^2 \; .
\ee
Eq. (42) is the Weierstrass equation whose solution
\be
z(t) = \frac{6}{\bt^3}{\cal P}(t-t_0)
\ee
is expressed through the Weierstrass function [47]. The time $t_0$ plays
the role of the escape time [36,37] and is defined from the initial
condition $z(0)=z_0$, which gives
\be
t_0 =\sqrt{\frac{3}{2\bt^3}}\int_{z_0}^\infty
\frac{dz}{\sqrt{z^3-z_0^3+\zeta}}\; .
\ee
The first equation from (40) takes the form
\be
\frac{dx^2}{dt^2} = \frac{3}{\bt^2}{\cal P}(t-t_0) x \; ,
\ee
which is the Lam\'e equation [48] whose degree is defined by the equality
$n(n+1)=3/\bt^2$, with $n>0$, which yields
\be
n = \frac{1}{2}\left ( \sqrt{1+\frac{12}{\bt^2}} - 1\right ) \; .
\ee
The solution to (46) can be expressed through the Lam\'e functions [48].

From the properties of the Weierstrass functions [47] it follows that the
axial motion is bounded from below by the minimal value
\be
z_{min} = \left ( z_0^3 -\zeta\right )^{1/3} \; .
\ee
When $t\ra t_0$, then
\be
x(t) \sim | t - t_0 |^{-1/2}\; , \qquad
z(t) \sim | t - t_0 |^{-2} \; .
\ee
The aspect ratio
$$
\frac{x(t)}{z(t)}\sim | t - t_0 |^{3/2} \ra 0
$$
shows that the atom trajectories are stretched in the axial direction.

The asymptotic expressions (49) give, of course, only a qualitative
understanding of the collimation process, since equations (41) and (44) are
valid only for the initial stage of atomic motion, when $x$ and $z$ are
small. To consider the dynamics of atoms for arbitrary times, we have to
solve Eqs. (40) numerically. We accomplished such a numerical investigation 
for the parameter $\bt=2$ and $\ep=1$. The initial positions of atoms are
taken in the center of a trap, with varying initial velocities
$|\dot{x}_0|\leq 1$ and $|\dot{z}_0|\leq 1$. The atomic trajectories in
the $x-z$ plane are presented in Figs. 1 to 3. The phase portraits for the
velocities
\be
v(t) \equiv \frac{dx}{dt}\; , \qquad
w(t)\equiv \frac{dz}{dt}
\ee
are shown in Figs. 4 to 6. These figures demonstrate that the bunch of
atomic trajectories is essentially squeezed in the radial direction
forming a well collimated beam.

\section{Constant Field}

The limiting case of a transverse field satisfying the second inequality
in (15) corresponds to a constant field with 
\be
h_x = const \; , \qquad h_y = const \; .
\ee
Then the effective frequency (14) is zero. The matrix $\hat A$ in (10),
with the elements (11), does not depend explicitly on time. Under $\sr$
fixed, the solution (19) becomes an exact solution to equation (10). For
the average force (30), we find
\be
\stackrel{\ra}{F} = \om_1^2\frac{(x+h_x)S_x^0+(y+h_y)S_y^0+\lbd zS_z^0}
{(x+h_x)^2+(y+h_y)^2+\lbd^2z^2} \left [ (x+h_x)\sse_x + (y+h_y)\sse_y +
\lbd^2 z\sse_z\right ] \; .
\ee

Note that for the initial spin polarization $S_x^0\neq 0,\; S_y^0=S_z^0=0$
the force (52) does not provide confinement, contrary to the average force
(32) related to the rotating field (31). However, confinement is not our
concern here. We wish to find a regime of semiconfinement, when the atoms
move predominantly in one direction.

Let us take for the initial spin polarization the initial condition (33).
Then the average force (52) becomes
\be
\stackrel{\ra}{F} = \om_0^2 \lbd ({\rm sgn} S) z
\frac{(x+h_x)\sse_x +(y+h_y)\sse_y + \lbd^2 z\sse_z}
{(x+h_x)^2+(y+h_y)^2+\lbd^2z^2} \; ,
\ee
where the notation (34) is used. For the trap form factor, we accept the
same expression (35). But, instead of (36), we now have
\be
f(\sr) =\frac{\vp(\sr)}{(x+h_x)^2+(y+h_y)^2+\lbd^2z^2} \; .
\ee
Passing to the dimensionless time measured in units of $\om^{-1}_0$, we
obtain as the equation of motion (26) the system of equations
$$
\frac{dx^2}{dt^2} = \lbd ({\rm sgn} S) ( x + h_x ) fz \; , \qquad
\frac{dy^2}{dt^2} = \lbd ({\rm sgn} S) ( y + h_y ) fz \; , 
$$
\be
\frac{dz^2}{dt^2} = \lbd^3 ({\rm sgn} S) fz^2 \; .
\ee
Eqs. (55), though are different form (37), possess the same invariance
property with respect to the transformations $\lbd\ra -\lbd,\; S\ra - S$
and $S\ra -S,\; z\ra -z$. Thence, we can again accept condition (38). The
integral of motion (39) exists for (55) only if $h_x=h_y$.

The system of equations (55) also provides the semiconfining regime of
motion. For example, at the initial stage, when $|\sr|\ll 1$ and
$f(\sr)\simeq 1$, assuming condition (38), we have
$$
\frac{dx^2}{dt^2} =\bt h_x z\; , \qquad
\frac{dz^2}{dt^2} = \bt^3 z^2 \; .
$$
The equation for the $y$--component is the same as for the variable $x$,
with the change of $h_x$ by $h_y$. The equation for $z$ coincides with
(41), thus, having the same solution (44), demonstrating that the atomic
motion is locked from below by the minimal value (48). When the time
approaches the escape time (45), then instead of (49), we now have
$$
x(t) \sim \ln | t - t_0|\; , \qquad z(t) \sim | t - t_0 |^{-2} \; ,
$$
which shows that, with the transverse constant field, the collimation is
even better than with the rotating field. At the late stage of the
process, when $|x/h_x|\gg 1$ and $|y/h_y|\gg 1$, the factors (54) and (36)
are asymptotically equal, and the evolution equations (55) and (37) become
equivalent.

In this way, the semiconfining regime of motion can be realized for
different magnetic fields satisfying condition (15), under the initial
spin polarization (33). This regime is clearly illustrated by numerical
calculations for the rotating transverse field. The numerical
investigation of motion for the constant transverse field will be given in
a separate paper. The semiconfining regime, we described, can be used for
creating well--collimated beams from atom lasers.

\vspace{5mm}

{\bf Acknowledgement}

\vspace{2mm}

We appreciate financial support from the University of Western Ontario,
Canada.

\newpage

{\Large{\bf Appendix A. Standard Form}}

\vspace{5mm} 

Equations (8) and (10) can be written in the standard form, as a system of
differential equations of first order,
$$
\frac{d\sr}{dt} = \stackrel{\ra}{p} \; , \qquad
\frac{d\stackrel{\ra}{p}}{dt} = \om_1^2 f_1 \; , \qquad
\frac{d\sS}{dt} = \om_2 f_2 \; ,
$$ with finite functions $f_1$ and $f_2$. By means of the notation
$$
t' \equiv \om_2 t \; , \qquad
\stackrel{\ra}{p}' =\frac{\stackrel{\ra}{p}}{\om_1} \; , \qquad
\ep \equiv \frac{\om_1}{\om_2} \; ,
$$
the above differential equations can be presented as
$$
\frac{d\sr}{dt'} = \ep \stackrel{\ra}{p}' \; , \qquad
\frac{d\stackrel{\ra}{p}'}{dt'} = \ep f_1 \; , \qquad
\frac{d\sS}{dt'} = f_2 \; .
$$
Condition (15) tells us that $\ep$ is a small parameter, $|\ep|\ll 1$.
Then from the latter system of equations it follows immediately that $\sr$
is a slow variable with respect to $\sS$. The same concerns the effective
velocity $\stackrel{\ra}{p}'$.

\vspace{1cm}

{\Large{\bf Appendix B. Gravitational Force}}

\vspace{5mm}

Since an atom has a mass, then, contrary to photon lasers, the existence
of gravity may influence the performance of atom lasers. The influence of
gravity is not essential for the confining regime, when we start with the
initial spin polarization $S_x^0<0,\; S_y^0=S_z^0=0$. Then, adding to the
evolution equation (8) the gravitational force $-mg\sse_z$, in the case of
the rotating field, we come to the equations
$$
\frac{d^2x}{dt^2} = -\frac{1}{2}|S_x^0|\om_1^2 x\; , \qquad
\frac{d^2z}{dt^2} = -\lbd^2|S_x^0|\om_1^2 z -\frac{g}{R_0} \; ,
$$ 
describing the motion of atoms inside the trap, where $|\sr|\ll 1$. We
do not write down the equation for $y$, which is similar to that for $x$.
These equations, as is evident, define an oscillatory motion in all three
dimensions. The sole thing that is changed, when taking account of
gravity, is the shift of the equilibrium position on the $z$--axis from
zero to $z_{eq}= -g(\lbd^2|S_x^0|\om_1^2R_0)^{-1}$. So that the center of
harmonic oscillations is shifted from the center of coordinates to the
point $\{ 0,0,z_{eq}\}$. Certainly, if the quadrupole field is switched
off, the atoms fall down because of gravity. But while the quadrupole
field is sufficient for confinement, the gravitational force does not
change principally the regime of motion corresponding to simple harmonic
oscillations. Recall that the constant transverse field does not provide
confinement in any case, so that the gravity is again of no importance.

In the case of the semiconfining regime, with the initial spin
polarization (33), the axis $z$ is an axis of the device, but not
necessary the vertical axis. Since the device can be oriented arbitrarily,
the gravitational force can also be directed along different axes. For
instance, assume that this force is $-mg\sse_x$. Then, for the rotating
transverse field, instead of (40), we have
$$
\frac{d^2x}{dt^2} = \frac{\bt}{2} f x z - \gm \; , \qquad
\frac{d^2z}{dt^2} = \bt^3 fz^2 \; ,
$$
where $\gm\equiv g/\om_0^2R_0$. To estimate the value of $\gm$, let us
take the parameters of the magnetic fields as in the quadrupole traps with
the rotating transverse field [43,44]. Then $\om_0\sim10^2-10^3$s$^{-1}$
and $R_0\sim 0.1-0.5$cm. With $g\approx 10^3$cm s$^{-2}$, this gives
$\gamma\sim 10^{-3}-1$. The equation for $z$ is the same as earlier and
describes the semiconfining motion along the $z$--axis. The motion in the
radial direction is also semiconfined, with the downward deviation of an
atomic beam because of gravity. Thus, the regime of motion does not change
principally.

If the device is oriented so that its $z$--axis is in the vertical
direction, then, instead of (40), we get
$$
\frac{d^2x}{dt^2} = \frac{\bt}{2} fxz\; , \qquad
\frac{d^2z}{dt^2} = \bt^3 fz^2 - \gamma\; .
$$
Integrating once the equation for $z$, we have
$$
\left (\frac{dz}{dt}\right )^2 = \frac{2}{3}\bt^3 \left [
z^3 - z_0^3 -\frac{3\gm}{\bt^3} (z - z_0) + \zeta\right ] \; .
$$
The latter equation can be reduced to the Weierstrass form
$$
\left (\frac{d{\cal P}}{dt}\right )^2 = 4{\cal P}^3 - g_2{\cal P} - g_3
$$
with the Weierstrass invariants
$$
g_2 =\frac{12\gm}{\bt^3}\; , \qquad
g_3 = \frac{2}{3}\bt^3z_0^3 - 2\gm z_0 - \dot{z}_0^2 \; .
$$
For $z$, then, we find
$$
z(t) = \frac{6}{\bt^3}{\cal P}(t-t_0) \; .
$$
When $g_2^3<27g_3^2$, then the motion along the $z$--axis is, as earlier,
semiconfined. But if $g_2^3>27g_3^2$, then the motion of atoms can become
confined depending on initial conditions for $z$. Recall again, that we
are not obliged to align the device axis along the gravitational force.
Therefore the semiconfining regime of motion can always be realized.

\vspace{5mm}

{\Large{\bf Appendix C. Initial Conditions}}

\vspace{3mm}

The evolution equations (1) and (2), we started with, are the equations
for the averages of the position and spin operators. The average of an
operator $\hat Q$ of an observable quantity is the scalar product $<\hat
Q>\equiv (\psi,\hat Q\psi)$, where $\psi=\psi(t)$ is the wave function
satisfying the time--dependent Schr\"odinger equation. The evolution
equation for such an average reads
$$
i\hbar\frac{\prt}{\prt t}<\hat Q> = < [\hat Q,\hat H]>\; ,
$$
where $\hat H$ is the Hamiltonian (see e.g. [49,50]). The initial state
$\psi_0=\psi(0)$ of the time--dependent Schr\"odinger equation is, in
general, an arbitrary function. Consequently, the initial value
$(\psi_0,\hat Q\psi_0)$ is also arbitrary. One tells that the initial
state can be prepared [50,51]. The evolution equation for the average
$\sS=\sS(t)=<\hat{\sS}>$ of the spin operator $\hat{\sS}$ is equation (2)
describing the spin precession from an arbitrary prepared initial value
$\sS_0=\sS(0)$, as is discussed, e.g., in Ref. [52]. In this way,
according to the basic principles of Quantum Mechanics, the initial state
can, in general, be chosen arbitrarily, resulting in an arbitrary value of
$\sS_0$. 

It is a different question how this or that particular state can be
realized in experiment. One distinguishes two limiting cases of preparing
initial conditions, {\it adiabatic} and {\it nonadiabatic} [53,54]. In the
first case, the time--dependent Schr\"odinger equation with a Hamiltonian
$\hat H(t)$ is ascribed an initial state $\psi_0$ that is an eigenstate of
$\hat H(0)$, i.e., $\hat H(0)\psi_0=E(0)\psi_0$. In the second case, a
chosen initial state $\psi_0$ is not an eigenstate of $\hat H(0)$. This
implies that the state $\psi_0$ could be prepared, for $t\leq 0$, as an
eigenstate of another Hamiltonian, say $\hat H_0$, and then, at the time
$t=0$, some external field is suddenly turned on, instantaneously
changing the Hamiltonian from $\hat H_0$ to $\hat H(0)$, so that $\hat
H_0\neq\hat H(0)$. These two types of initial conditions correspond to
different physical realizations of an {\it adiabatically slow} changing
field and of a {\it sudden switch on} of an external field [53,54]. Both
these cases can, in principle, be realized in experiment. Of course, when
one aims at trapping atoms, the adiabatic motion is necessary [55], with
adiabatic initial conditions. But this is not mandatory when one's aim is
to obtain a specific nonconfined regime. Then one should try nonadiabatic
initial conditions. Our conditions (33) for the initial spin polarization
are such nonadiabatic conditions providing the semiconfining regime of
motion which can be used for creating well--collimated beams of atom
lasers. This kind of initial conditions can be realized in practice in
different ways. For example, one can confine atoms in a trap with a
vertical bias field [45], where all atoms become spin--polarized along the
$z$--axis. After this, the bias field in the $z$--direction is suddenly
switched off and, at the same time, a transverse bias field, as in
experiments [43,44], is suddenly switched on. Another possibility could be
to prepare spin--polarized atoms in one trap and to suddenly load them
into another trap with the desired field configuration. We are not going
to plunge into technical details of preparing such initial conditions for
realizing the semiconfined motion. This is neither our aim nor our
speciality. To do this is the challenge for experimentalists. We think
that this semiconfined motion can be realized, since there are no
principal theoretical obstacles for it.

\newpage

\begin{center}

{\bf Figure Captions}

\end{center}

{\bf Fig. 1.} The trajectories of atoms at the initial stage of the
collimation process, $0\leq t\leq 5$, for a trap with $L=100$: (a)
isotropic initial conditions, $|\dot{x}_0|\leq 1$, $|\dot{z}_0|\leq 1$;
(b) slightly anisotropic initial velocities, $|\dot{x}_0|\leq 0.25$,
$|\dot{z}_0|\leq 1$.

\vspace{5mm}

{\bf Fig. 2.} Atomic trajectories for the time $0\leq t\leq 20$ for two
difefrent traps: (a) $L=10$; (b) $L=100$.

\vspace{5mm}

{\bf Fig. 3.} Atomic trajectories for long times, $0\leq t\leq 100$, for a
trap with $L=100$.

\vspace{5mm}

{\bf Fig. 4.} The velocities of atoms in the radial, $v(t)$, and axial,
$w(t)$, directions, with initial velocities $|\dot{x}_0|\leq 1$ and
$|\dot{z}_0|\leq 1$ for two cases: (a) $0\leq t\leq 100,\; L=100$, (b)
$0\leq t\leq 200,\; L=10000$.

\vspace{5mm}

{\bf Fig. 5.} Atomic velocities for the initial conditions
$|\dot{x}_0|\leq 0.1$ and $|\dot{z}_0|\leq 0.1$ for the following cases:
(a) $0\leq t\leq 100,\; L=100$; (b) $0\leq t\leq 200,\; L=1000$.

\vspace{5mm}

{\bf Fig. 6.} The velocities of atoms for the initial conditions
$|\dot{x}_0|\leq 0.1$ and
$|\dot{z}_0|\leq 0.1$ for $L=10000$, at different stages: (a) initial
stage, $0\leq t\leq 10$; (b) long times, $0\leq t\leq 200$. For $t>200$,
the picture practically does not change.

\end{document}